\documentclass [twocolumn,preprintnumbers,showpacs,amssymb] {revtex4}

\usepackage{epsfig,graphicx}
\usepackage{dcolumn}
\usepackage{bm}
\usepackage{times}

\draft

\begin {document}

\title{Spatial prisoner's dilemma game with volunteering in Newman-Watts small-world networks}

\author{Zhi-Xi Wu, Xin-Jian Xu, Yong Chen\footnote {Email address: ychen@lzu.edu.cn}, and Ying-Hai Wang\footnote {Email address: yhwang@lzu.edu.cn}}

\address{Institute of Theoretical Physics, Lanzhou University, Lanzhou Gansu 730000, China}

\begin {abstract}
A modified spatial prisoner's dilemma game with voluntary
participation in Newman-Watts small-world networks is studied.
Some reasonable ingredients are introduced to the game
evolutionary dynamics: each agent in the network is a pure
strategist and can only take one of three strategies (\emph
{cooperator}, \emph {defector}, and \emph {loner}); its
strategical transformation is associated with both the number of
strategical states and the magnitude of average profits, which are
adopted and acquired by its coplayers in the previous round of
play; a stochastic strategy mutation is applied when it gets into
the trouble of \emph {local commons} that the agent and its
neighbors are in the same state and get the same average payoffs.
In the case of very low temptation to defect, it is found that
agents are willing to participate in the game in typical
small-world region and intensive collective oscillations arise in
more random region.
\end {abstract}


\maketitle

There has been a long history of studying complex behaviors
qualitatively of biological, ecological, social and economic
systems using special game models. After the prisoner's dilemma
game (PDG) was first applied by Neumann and Morgenstern \cite
{Neumann} to study economic behavior, great developments have been
made by a lot of subsequent studies. Recently, more and more
attentions have been focused on the applications of the PDG in the
fields of biology \cite {Wahl}, economy \cite {Fehr}, ecology
\cite {Mesterton-Gibbons}, and other domains \cite {Huberman}.
Game theory and evolutionary theory provide a powerful metaphor
for simulating the interactions of individuals in these systems
\cite {Maynard}.

Most realistic systems can be regarded as composing of a large
number of individuals with simple local interactions. For example,
human beings are limited in territory and interact more frequently
with their neighbors than those far away. Therefore, the spatial
structure may greatly affect their activities. Since Axelrod
\cite{Axelrod} suggested ideas of the PDG on a lattice, spatial
prisoner's dilemma games (SPDG) have been extensively explored in
various kinds of network models in the past few years, including
regular lattices \cite {Nowak, Szabo_0, Vainstein}, random regular
graphs \cite {Szabo_1}, random networks with fixed mean degree
distribution \cite {Ebel}, small-world networks \cite {Szabo_2,
Kim, Abramson} and real-world acquaintance networks \cite {Holme},
etc. In the general SPDG, each agent can take one of two
strategies (or states): \emph {cooperator} ($C$) and \emph
{defector} ($D$). There are four possible combinations: ($C$,
$C$), ($C$, $D$), ($D$, $C$) and ($D$, $D$), which get payoffs
($r$, $r$), ($s$, $t$), ($t$, $s$), and ($p$, $p$), respectively.
The parameters satisfy the conditions $t>r>p>s$ and $2r>t+s$, so
that lead to a so-called dilemma situation where mutual trust and
cooperation is beneficial in a long perspective but egoism and
guile can produce big short-term profit. Agents update their
states by imitating the strategy of the wealthiest among their
neighborhoods in subsequent plays. The system is easy to get into
an absorbing state: all agents are $D$ for large values of $t$,
which is known as the tragedy of the \emph{commons }\cite
{Hardin}.

Recently, Szab\'o \emph{et al.} \cite {Szabo_0, Szabo_1, Szabo_2}
developed the SPDG with voluntary participation, in which agents
can take one of three possible strategies, \emph {cooperator},
\emph {defector} and \emph {loner} ($L$). \emph {Cooperators} and
\emph {defectors} are interested in taking part in the game and
the payoffs for their encounters are assigned as before. \emph
{Loners} do not participate in the game temporarily and get the
same small fixed income $\sigma$ ($\sigma<r<t$) as their
neighbors. Thus the payoff matrix can be tabulated as
\begin {eqnarray}
\begin {array}{c c c c}
&{\bf C} & {\bf D} & {\bf L}\\ {\bf C} & r & s & \sigma \\
 {\bf D} & t & p & \sigma \\  {\bf L} & \sigma & \sigma &
\sigma.
\end {array}
\end {eqnarray}
Each element in the matrix denotes the corresponding payoff of an
agent adopting the strategy of the left and encountering an agent
performing the strategy of the above. In the volunteers version,
the three strategies can coexist by cyclic dominance ($D$ invades
$C$ invades $L$ invades $D$), which efficiently avoid the system
getting into a frozen state.

In this Brief Report, we study the SPDG with voluntary
participation in the Newman-Watts (NW) network, which is a typical
small-world model constructed as follows: starting with a
two-dimensional lattice with periodic boundary conditions; each
agent locates on the lattice and links with its four nearest
neighbors; for every agent, with probability $Q$, we add a long
range link for each its four links to a random selected agent from
the whole system with duplicate links forbidden; then a NW network
is realized (see Ref. \cite {Newman_1} for details). The
structural characteristics of social communities, namely, high
clustering and small diameter, can be well described by this
small-world graph. A round of play consists of the encounters of
all agents with their nearest neighbors. Following Ref. \cite
{Kim}, the payoffs earned by the agents are calculated as average
and not accumulated from round to round. To start the next round,
agents are allowed to inspect the profits collected by their
neighbors and adjust their strategies.

We argue that the ingredients for agents changing their states
mainly come from two aspects: (i) For the sake of pursuing higher
profits, agents have a trend to follow the successful agents who
get higher payoffs, i.e., \lq\lq successful\rq\rq strategies are
imitated. We figure that $i$th agent adopts the strategy of its
arbitrary neighbor $j$ with a probability
\begin {equation}
\gamma_{ij}=\frac{g_{j}}{\sum_{k\in\Omega_{i}} g_{k}}\label{rule},
\end {equation}
where $g_{j}$ denotes the average profit earned by $j$th agent and
$\Omega_{i}$ is the community composing of the nearest neighbors
of $i$ and itself; (ii) When one agent and its neighbors are in
the same state and get the same average payoffs, it has a
spontaneous willing to make some mutations. We propose that the
agents getting into the above case make spontaneous alterations
with a probability depending on the elements of the payoff matrix.
If the agent under consideration is $C$, in the next round, the
probabilities for its changing to $C$, $D$, or $L$ are
$r/(r+t+\sigma)$, $t/(r+t+\sigma)$, and $\sigma/(r+t+\sigma)$
respectively; if the agent is $D$, the probabilities for its
changing to $C$, $D$, or $L$ are $s/(s+p+\sigma)$,
$p/(s+p+\sigma)$, and $\sigma/(s+p+\sigma)$ respectively; and if
the agent is $L$, the probabilities of its changing to $C$, $D$,
or $L$ are the same value and equal to $1/3$. This spontaneous
mutational mechanism not only efficiently avoids the system
getting into a frozen state but also sufficiently describes the
agents' flexibility.

\begin {figure}
\centerline{\epsfxsize=8cm \epsffile{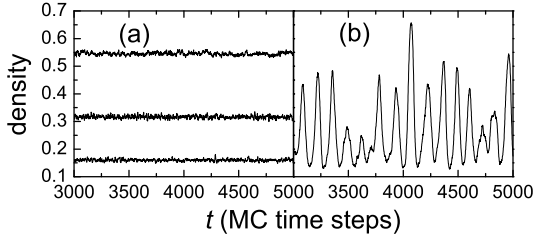}} \caption{The
evolution of the density of \emph {defectors} ($\rho_{D}$) with
varied values of (RTQ, $Q$) under the equilibrium state: (a) form
top to bottom, the curves correspond to ($0.02$, $0.1$), ($0.56$,
$0.1$), ($0.56$, $0.5$) respectively; and (b) ($0.02$, $0.5$).}
\label {fig1}
\end {figure}

\begin {figure}
\centerline{\epsfxsize=8cm \epsffile{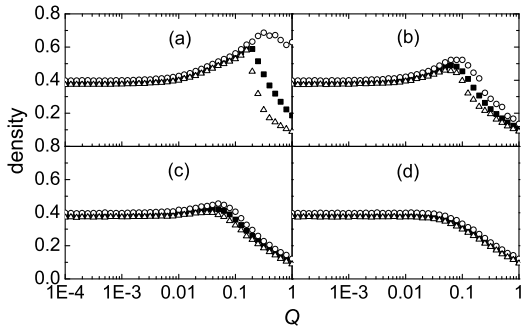}} \caption{MC data
of the density of defectors as a function of the network's
structure parameter $Q$ under different values of RTQ: 0.02 (a),
0.1 (b), 0.2 (c) and 0.8 (d). Closed squares represent the average
density of defectors; open circles and triangles show their
maximal and minimal values due to oscillation.} \label {fig2}
\end {figure}

\begin {figure}
\centerline{\epsfxsize=8cm \epsffile{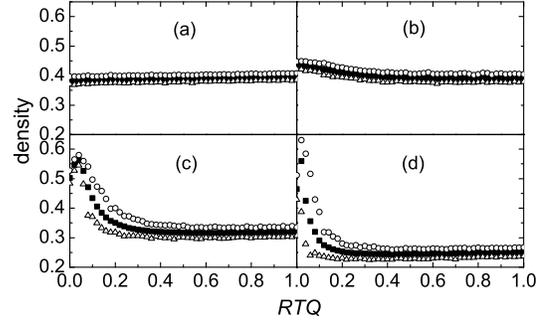}} \caption{MC data
of the density of defectors as a function of RTQ under different
values of the network's structure parameter Q: 0.001 (a), 0.1 (b),
0.5 (c) and 1.0 (d). The symbols as shown in Fig. \ref
{fig2}.}\label {fig3}
\end {figure}

\begin {figure}
\centerline{\epsfxsize=6cm \epsffile{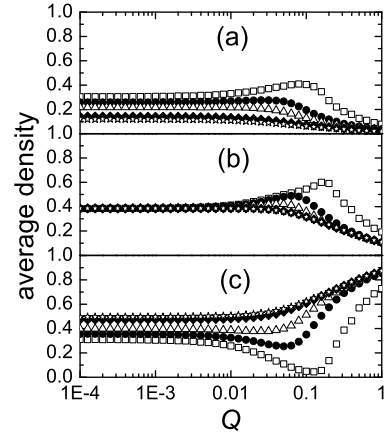}} \caption{The
density of \emph {cooperators} (a), \emph{ defectors} (b) and
\emph{loners} (c) vs the network's structure parameter $Q$ under
different values of RTQ. The symbols of open squares, closed
circles, open triangles, closed diamonds and open stars correspond
to the value of RTQ: $0.02$, $0.1$, $0.2$, $0.56$, and $0.8$,
respectively.}\label {fig4}
\end {figure}

\begin {figure}
\centerline{\epsfxsize=6cm \epsffile{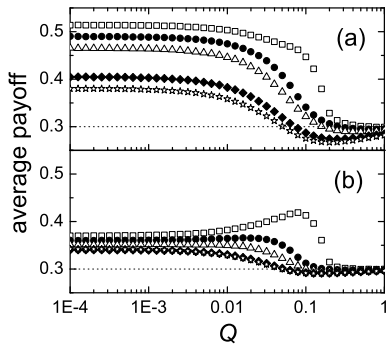}} \caption{Average
payoffs of \emph {cooperators} (a) and \emph {defectors} (b) vs
the network's structure parameter $Q$ under different values of
RTQ: $0.02$, $0.1$, $0.2$, $0.56$, and $0.8$. The symbols are the
same as shown in Fig. \ref {fig4}, and the dotted line indicates
the fixed average payoff of \emph {loners}.}\label {fig5}
\end {figure}

Our analysis of the model is based on systematic Monte Carlo (MC)
simulations performed in different NW networks with the total size
of $200 \times 200$ populations. The three strategies are assigned
randomly to the agents with probability $1/3$ initially. For
convenience, following Refs. \cite {Szabo_1, Nowak, Abramson,
Kim}, we set $s=p=0$, $r=1$, $\sigma=0.3$, and $1<t<2$. We define
$t-r$ as the relative temptation quantity (shortly RTQ) reflecting
the extent of the temptation and cursorily partition the networks
into three regions: lattice, small-world and random graphs
corresponding to the variational range of $Q$: $(0.0001, 0.001)$,
$(0.001, 0.3)$ and $(0.3, 1)$ respectively. We iterate the rules
of the model with parallel updating. The total sampling times are
$5000$ MC steps. After appropriate relaxation times the system
stabilizes in dynamical equilibrium characterized by their
densities of $\rho_{C}$, $\rho_{D}$, $\rho_{L}$ and average
payoffs $P_{C}$, $P_{D}$, $P_{L}$. According to the previous
assumption, it is easy to know that $P_{L}$ is always equals to
$\sigma$. All the results are averaged over the realizations of
ten networks.

The main features of the steady-state phase diagram can be
summarized as follows. All three states coexist and coevolve
steadily in equilibrium state. For large values of $Q$ with very
small values of RTQ, strong global oscillations arise, which is
similar to the phenomena studied in Ref. \cite {Szabo_2} for high
temptation to defect. The bifurcation of $\rho_{D}$ for large
values of temptation studied in Refs. \cite {Szabo_1, Szabo_2},
however, does not arise in our model. For small values of $Q$ with
arbitrary values of RTQ or large values of RTQ with arbitrary
values of $Q$, the stationary state is characterized by a weak
global oscillation where the amplitude of fluctuation is
significantly less than the corresponding average value. As a
distinct view, in Fig. \ref{fig1}, the last $2000$ steps'
evolution of $\rho_{D}$ under values of $Q$ ($0.1$ and $0.5$) and
RTQ ($0.02$ and $0.56$) has been tracked (the evolution of
$\rho_{C}$ and $\rho_{L}$ are similar to $\rho_{D}$); the average
values of $\rho_D$ and the corresponding maximum and minimum
deviation in the steady state are also reported in Fig. \ref{fig2}
for fixed values of RTQ $(0.02, 0.1, 0.2, 0.8)$ with varied values
of $Q$ $\in (0.0001 \thicksim 1.0)$ and in Fig. \ref {fig3} for
fixed values of $Q$ $(0.001, 0.1, 0.5, 1.0)$ with varied values of
RTQ $\in (0.0\thicksim 1.0)$. These phenomena can be explained as
follows.

During the process of the evolution, \emph {defectors} can not
form stable large clusters, of which the inner agents would get
zero profit and possess the same state as their neighbors
(\emph{local commons}). According to the evolutionary rules, they
will try to throw off embarrassment by changing their strategies.
Namely, the easy formation of clusters of $D$ will make the agents
self-adapt frequently in their communities, and then confine the
fluctuation of $ \rho_{D}$ in a narrow range [see Fig.
\ref{fig1}(a), Fig. \ref{fig2}(b), Fig. \ref{fig2}(c), Fig.
\ref{fig2}(d), Fig. \ref{fig3}(a), Fig. \ref{fig3}(b)]. There are
two factors favoring the forming of clusters of \emph{defectors}:
the high temptation to defect (large values of RTQ) and the well
clustered structure of the agents (small values of $Q$), which
would strengthen the adoption and the imitation of strategy $D$
greatly. Therefore, in our model, high temptation to defect will
only give rise to steady oscillation of the system rather than
result in the bifurcation phenomena studied in Refs. \cite
{Szabo_1, Szabo_2}. While for poorly clustered agents (large
values of $Q$) with low temptation, the formation of large
clusters of \emph{defectors} is reasonably difficult, which would
slow down the evolutionary velocity of the whole system and
guarantee the growth (decline) of $\rho_{D}$ lasting for a long
time, and consequently broaden the fluctuant amplitude (see Fig.
\ref{fig1}(b), Fig. \ref{fig2}(a), Fig. \ref{fig3}(c) and Fig.
\ref{fig3}(d).

In addition, in the lattice region, $\rho_{D}$ keeps a steady
level for any values of RTQ [see Fig. \ref{fig2}, Fig.
\ref{fig3}(a) and Fig. \ref{fig4}(b)]. It is also a result of the
fast self-adaptation of the agents. With the increasing of RTQ,
agents of $C$ are easy to change to $D$ for high temptation, and
then again change to $L$ because clusters of \emph{defectors} are
extremely unstable and can not survive a long time. The decrease
of $\rho_C$ nearly results in the increasing of $\rho_L$ [see Fig.
\ref{fig4}(a) and Fig. \ref{fig4}(c)]. In this region, the fast
self-adaptation of the agents also leads to the case that the
neighbors of \emph{defectors} would include other types of agents
in most time during the evolution, which gives rise to larger
values of $P_D$ than $P_L$. By comparison, in Refs. \cite{Szabo_1,
Szabo_2}, very big clusters of \emph{defectors} can survive a long
time during the evolution and most agents would get only the zero
payoff resulting in lower average payoffs of the \emph{defectors}
than the \emph{loners}. It is obvious that the differences in the
evolutionary dynamics of the game give rise to the distinct
results. It is worth mentioning that the present model is also
different from the cyclic spatial games studied in Ref. \cite
{Szabo_3} where the dynamics evolution is governed by a strictly
cyclic dominance, i.e., $\emph {rock}$ dominates $\emph
{scissors}$ dominates $\emph {paper}$ dominates $\emph {rock}$.
While in our model, any two types of the three strategies can
transform each other in particular case. As a result of the
difference in evolutionary dynamics, the phase transitions
phenomena studied in Ref. \cite {Szabo_3} for $\emph
{rock-scissors-paper}$ games do not arise in our model.

Another interesting feature of the equilibrium phase diagram is
that in the vicinity of $Q=0.1$ where the NW networks possess
notable small-world effect, namely, large clustering and small
diameter at the same time, agents are willing to participation in
the game in the case of very low temptation to defect. To view in
detail, in Fig.\ref {fig4} and Fig.\ref {fig5}, we plot the
average density and corresponding average payoffs of $C$-$D$-$L$
vs the small-world parameter $Q$ under different values of RTQ
respectively. For very low temptation to defect (e.g. RTQ$=0.02$),
the evolutionary curve of $\rho_{L}$ decreases slowly with the
increasing of $Q$ and reaches a minimum at certain culminating
point. As $Q$ increases over this point, $\rho_{L}$ ascends
rapidly (see Fig. \ref{fig4}). We conclude that two factors, the
very low temptation to defect and the small-world property of the
network, are beneficial for the spreading of $C$ in the system,
which then stimulates more and more agents to take part in the
game. In the case of more random networks $(Q\rightarrow1)$, the
evolutionary results of the game are qualitatively the same as
Refs. \cite{Szabo_1, Szabo_2}, i.e., the majority of members in
the system are \emph{loners} and the values of $P_{C}$ and $P_{D}$
get closed to the fixed value $\sigma$.

In summary, we have studied the SPDG with voluntary participation
in NW small-world networks. To model the realistic social systems,
some reasonable ingredients are introduced to the evolutionary
dynamics: each agent in the networks is a pure strategist and can
only take one of three strategies (\emph {C, D, L}); its
strategical transformation is associated with both the number of
strategical states and the magnitude of average profits, which are
adopted and acquired by its coplayers in the previous round of
play. To model initiative and flexibility, a stochastic strategy
change is applied when the agents get into the condition of \emph
{local commons}. The agents self-adapt and self-organize into
dynamical equilibrium after a short transient. When the agents are
well structured (the cases of small values of $Q$), they can
steadily coexist and coevolve. On the other hand, for high
temptation or more random networks, \emph {loners} dominate the
network. Especially, in the case of very low temptation to defect,
it is found that agents are willing to participate in the game in
typical small-world region and intensive collective oscillations
arise in more random region.

%\bigskip

The authors thank Prof. Hong Zhao for interesting discussions.
This work was partly supported by the National Natural Science
Foundation of China under Grant No. 10305005 and the Natural
Science Foundation of Gansu Province under Grant No.
ZS011-A25-004-2.

\end {document}